\newcommand{\manuscript}{3}

\newcommand{\be}{\begin{equation}}
\newcommand{\ee}{\end{equation}}
\newcommand{\ba}{\begin{eqnarray}}
\newcommand{\ea}{\end{eqnarray}}

\ifnum\manuscript=2
\documentclass[11pt, letter]{article}

\newcommand{\shorttitle}[1]{}
\newcommand{\shortauthors}[1]{}
\newcommand{\received}[1]{}
\newcommand{\altaffiltext}[2]{#1 #2}
\newcommand{\altaffilmark}[1]{\ensuremath{^{\mathrm{#1}}}}

\newcommand{\acknowledgements}[1]{ }
\renewcommand{\author}[1]{#1}
\renewcommand{\title}[1]{{\centering {\Large \bf #1}\\ }}

\else\ifnum\manuscript=1
\documentclass[12pt,letterpaper,preprint,flushrt]{aastex}
\else\ifnum\manuscript=3
\documentclass{emulateapj}
\usepackage[citebordercolor={0 .5 .5}]{hyperref}

\else
\documentclass{emulateapj}
\fi
\usepackage{graphicx}        % AASTeX preferred
\usepackage{bm}              % boldface math
\usepackage{natbib}
\usepackage{mathrsfs}
\newcommand{\commentx}[1]{}

\renewcommand{\vec}[1]{\mbox{\boldmath$#1$}} % Bold
\newcommand{\ra}[3]   % right ascension
   {\makebox[1.5em][r]{#1}\makebox[1.5em][r]{#2} \makebox[2em][r]{#3}}

\newcommand{\hms}[3]  % Write HMS in ApJ style.
   {${#1}^{\mathrm{h}}{#2}^{\mathrm{m}}{#3}^{\mathrm{s}}$}

\newcommand{\hmin}[2]  % Write HM in ApJ style.
   {\ensuremath{{#1}^{\mathrm{h}}{#2}^{\mathrm{m}}}}
   
\newcommand{\hours}[1]  % Write H in ApJ style.
   {\ensuremath{{#1}^{\mathrm{h}}}}

\newcommand{\dms}[3]  % Write DMS in ApJ style.
   {\ensuremath{{#1}\degree{#2}\arcminute{#3}\arcsecond}}

\newcommand{\dm}[2]  % Write DM in ApJ style.
   {\ensuremath{{#1}\degree{#2}\arcminute}}

\newcommand{\ukcmb}  % microKelvin_cmb
           {\ensuremath{\micro \kelvin_\mathrm{cmb}}}

\newcommand{\uk}  % microKelvin
           {\ensuremath{\micro \kelvin}}

\newcommand{\fdeg} % fractional degrees
           {\hbox{$.\!\!^{\circ}$}}

\usepackage{color}

\usepackage[mediumspace,Gray,squaren]{SIunits}
\usepackage{rotating}
\usepackage{graphicx}
\usepackage{amsmath}
\shorttitle{ACT x Planck}

\begin{document}
\title{The Atacama Cosmology Telescope:  Cross Correlation with {\it Planck} maps}
%%%%%%%%%%%%%%%%%%%%%%%%%%%%%%%%%%%%%%%%%%%%%%%%%%%%%%%%%%%%%%%%%%
%%%%%%%%%
% WARNING: This LaTeX block was generated automatically by authors.py
% Do not change by hand: your changes will be lost.

%\input{psAuthors.tex}
%%%%%%%%%%%%%%%%%%%%%%%%%%%%%%%%%%%%%%%%%%%%%%%%%%%%%%%%%%%%%%%%%%%%%%%%%%%
% WARNING: This LaTeX block was generated automatically by authors_das.py
% Do not change by hand: your changes will be lost.

\author{
Thibaut Louis\altaffilmark{1},
Graeme~E.~Addison\altaffilmark{2},
Matthew~Hasselfield\altaffilmark{3},
J~Richard~Bond\altaffilmark{4},
Erminia~Calabrese\altaffilmark{1},
Sudeep~Das\altaffilmark{5},
Mark~J.~Devlin\altaffilmark{6},
Joanna~Dunkley\altaffilmark{1},
Rolando~D\"{u}nner\altaffilmark{7},
Megan~Gralla\altaffilmark{8},
 Amir Hajian\altaffilmark{4},
Adam~D.~Hincks\altaffilmark{2},
Ren\'ee~Hlozek\altaffilmark{3},
Kevin Huffenberger\altaffilmark{10}, 
Leopoldo~Infante\altaffilmark{7}, 
Arthur~Kosowsky\altaffilmark{11}, 
Tobias A. Marriage\altaffilmark{8,9}, 
 Kavilan Moodley\altaffilmark{12},
Sigurd N${\ae}$ss\altaffilmark{1},
Michael~D.~Niemack\altaffilmark{ 9,13,14},
Michael R. Nolta\altaffilmark{4},
Lyman~A.~Page\altaffilmark{9}, 
Bruce~Partridge\altaffilmark{16}, 
Neelima Sehgal\altaffilmark{17}, 
Jonathan~L.~Sievers\altaffilmark{4,15,9},
David~N.~Spergel\altaffilmark{3},
Suzanne T. Staggs\altaffilmark{9},
Benjamin Z. Walter\altaffilmark{16},
Edward J. Wollack\altaffilmark{18}.
}
\altaffiltext{1}{Department of Astrophysics, Oxford University, Oxford OX1 3RH, UK}
\altaffiltext{2}{Department of Physics and Astronomy, University of British Columbia, Vancouver, BC V6T 1Z4, Canada}
\altaffiltext{3}{Department of Astrophysical Sciences, Peyton Hall, Princeton University, Princeton, NJ 08544, USA}
\altaffiltext{4}{Canadian Institute for Theoretical Astrophysics, University of Toronto, Toronto, ON M5S 3H8, Canada}
\altaffiltext{5}{Argonne National Laboratory, 9700 S. Cass Ave., Lemont, IL 60439}
\altaffiltext{6}{Department of Physics and Astronomy, University of Pennsylvania, 209 South 33rd Street, Philadelphia, PA 19104, USA}
\altaffiltext{7}{Departamento de Astronom{\'{i}}a y Astrof{\'{i}}sica, Facultad de F{\'{i}}sica, Pontificia Universidad Cat\'{o}lica de Chile, Casilla 306, Santiago 22, Chile}
\altaffiltext{8}{Dept. of Physics and Astronomy, The Johns Hopkins University, 3400 N. Charles St., Baltimore, MD 21218-2686}
\altaffiltext{9}{Joseph Henry Laboratories of Physics, Jadwin Hall, Princeton University, Princeton, NJ 08544, USA}  
\altaffiltext{10}{Department of Physics, Florida State University, Keen Physics Building, 77 Chieftan Way, Tallahassee, Florida, U.S.A.}
\altaffiltext{11}{Department of Physics and Astronomy, University of Pittsburgh, Pittsburgh, PA, USA 15260}
\altaffiltext{12}{Astrophysics and Cosmology Research Unit, School of Mathematical Sciences, University of KwaZulu-Natal, Durban, 4041, South Africa}
\altaffiltext{13}{NIST Quantum Devices Group, 325 Broadway Mailcode 817.03, Boulder, CO 80305, USA}
\altaffiltext{14}{Department of Physics, Cornell University, Ithaca, NY, USA 14853}
\altaffiltext{15}{Astrophysics and Cosmology Research Unit, School of Mathematical Sciences, University of KwaZulu-Natal, Durban, 4041, South Africa}
\altaffiltext{16}{Department of Physics and Astronomy, Haverford College, Haverford, PA 19041, USA}
\altaffiltext{17}{Dept. of Physics and Astronomy, Stony Brook University, Stony Brook, NY 11794-3800, USA}
\altaffiltext{18}{NASA/Goddard Space Flight Center, Greenbelt, MD 20771, USA}

\begin{abstract}
We present the temperature power spectrum of the Cosmic Microwave Background obtained by cross-correlating maps from the Atacama Cosmology Telescope (ACT) at 148 and 218 GHz  with maps from the {\it Planck} satellite at 143 and 217 GHz, in two overlapping regions covering 592 square degrees. We find excellent agreement between the two datasets at both frequencies, quantified using the variance of the residuals between the ACT power spectra and the ACT$\times${\it Planck} cross-spectra. We use these cross-correlations to calibrate the ACT data at 148 and 218 GHz, to 0.7\% and 2\% precision respectively. We find no evidence for anisotropy in the calibration parameter.  We compare the {\it Planck} 353 GHz power spectrum with the measured amplitudes of dust and cosmic infrared background (CIB) of ACT data at 148 and 218 GHz. We also  compare planet and point source measurements from the two experiments.
%We also assess the consistency of the two data sets by performing a $\chi^{2}$ test using an estimate of the variance of the residuals, $ r= C^{A \times A } -C^{A \times P}$ , that includes noise estimation and beam error propagation, but no cosmic variance.
\end{abstract}

%%%%%%%%%%%%%%%%%%%%%%%%%%%%%%%%%%%%%%%%%%%%%%%%%%%%%%%
%%%%%%%%%
\section{INTRODUCTION}

Recent measurements of the Cosmic Microwave Background (CMB) power spectrum by the Atacama Cosmology Telescope (\citealt{Das:2013zf}, hereafter D13), South Pole Telescope \citep{Story:2012wx}, and the {\it Planck} satellite  \citep{Ade:2013ktc,Planck:2013kta} provide a precise view of CMB anisotropies over a wide range of scales (2 $< \ell <$ 4000), extending the measurements by the {\it WMAP} satellite \citep{Hinshaw:2003ex,Bennett:2012zja} and earlier observations, and demonstrating exquisite agreement with the $\Lambda$CDM model \citep[e.g.,][]{ Spergel:2003cb, Hinshaw:2012aka, Sievers:2013ica, Story:2012wx, Ade:2013zuv}. 
%However some tensions between the different CMB data sets have been noted, and it is crucial to quantify their statistical significance.

We present a measurement of the cross-correlation of CMB temperature anisotropies at 148~GHz and 218~GHz, from the ACT data acquired during the 2008, 2009 and 2010 observing seasons, with the publicly released {\it Planck} maps at 143~GHz and 217~GHz. The overlap of the two experiments allows us to test their consistency. The ACT data used in the previous analysis were calibrated to {\it WMAP} data by matching the ACT$\times${\it WMAP} cross-spectrum to the ACT power spectrum \citep[at 2\% precision for ACT 148 GHz;] []{Hajian:2010fj}. The lower level of noise, higher resolution, and closer match in frequency of the {\it Planck} satellite data now enable a more precise calibration.

The paper is organized as follows. In Section \ref{data} we describe the observations, beam transfer function and mask used in the cross-correlation analysis. In Section \ref{results} we show the ACT$ \times${\it Planck} power spectra for each ACT season and each frequency, and use a simple likelihood to assess the consistency with the ACT power spectra and to compute a best-fit calibration factor. In Section \ref{dust}, we compute new estimates of the Galactic cirrus contamination and Cosmic Infrared Background (CIB) fluctuations in ACT maps using the 353~GHz {\it Planck} data. In Section \ref{Iso} we test the isotropy of the best fit calibration, by comparing the ACT two-dimensional power spectra to the ACT $\times$ {\it Planck} two-dimensional spectra. Lastly, we compare planet temperature measurements and flux density measurements of compact sources from the two experiments in Section \ref{planet} and Section \ref{sources}, respectively. We conclude in Section \ref{conclude}.

\begin{figure*}
\begin{center}
\includegraphics[width=19cm]{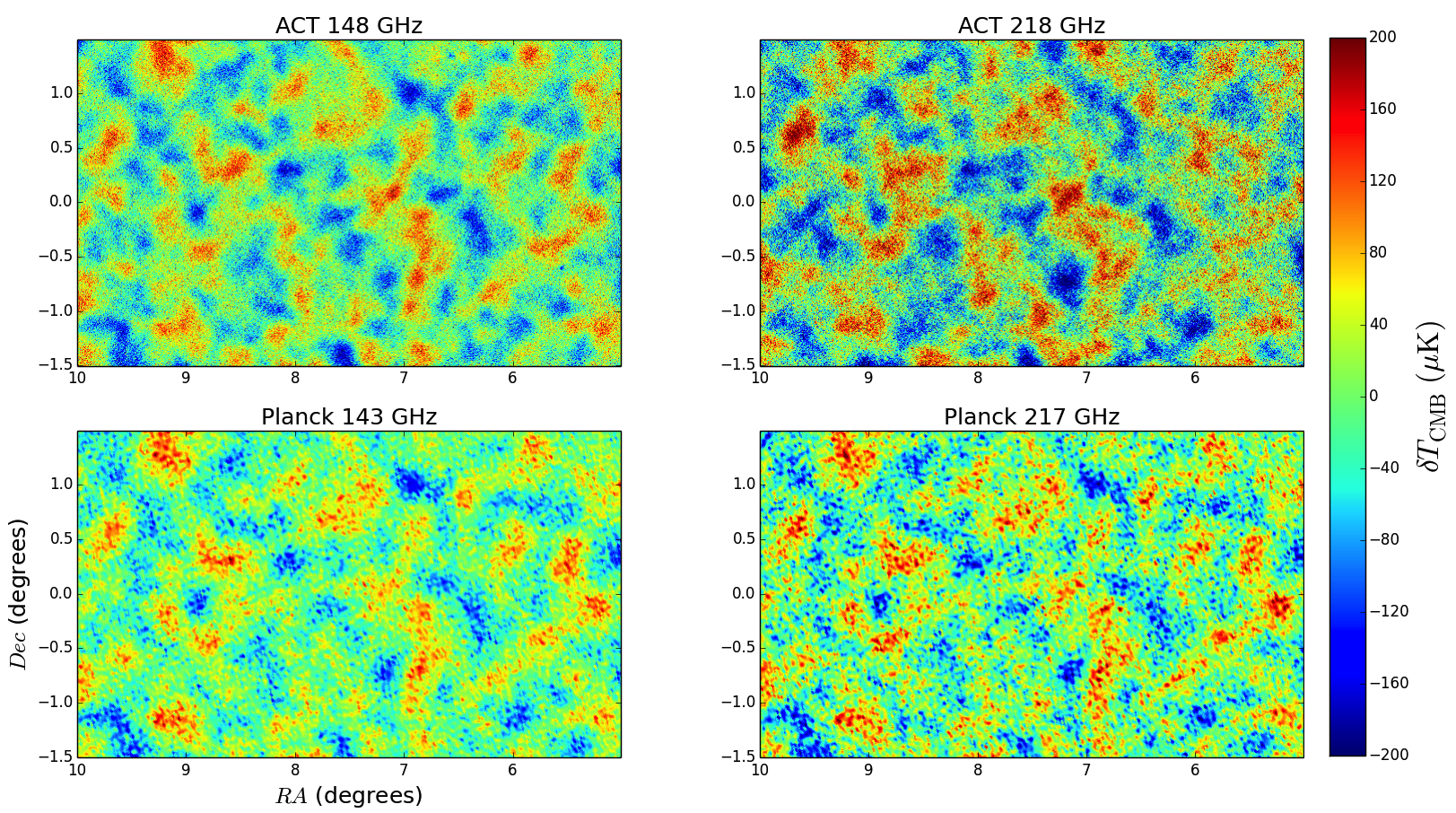}\\
\end{center}
\caption{Comparison of ACT ({\it top}) and {\it Planck} ({\it bottom}) maps for a 15 deg$^{2}$ patch in the ACT Equatorial region. The maps are the inverse variance weighted combination of all ACT data at  148~GHz ({\it left}) and 218~GHz ({\it right}) and all {\it Planck} data at 143~GHz and 217~GHz. All maps have been filtered with a high pass filter (for modes on scales: $\ell<500$). Artifacts of the HEALpix pixelization are seen in the {\it Planck} maps. The agreement is visually excellent. }
\label{Fig:MapComparison}
\end{figure*}

In this paper we take as a baseline the maps presented in D13. These maps were calibrated to {\it WMAP} as in \cite{Hajian:2010fj}. Recalibration factors $y$ quoted in the following are values by which one multiplies the ACT maps. That is, a recalibration factor of 1.01 means multiply the ACT map by 1.01 and increase the power spectrum by $(1.01)^{2}$. Note that the observed map calibration factor between {\it Planck} and {\it WMAP} maps is 0.985 ({\it Planck} is lower than {\it WMAP}; \cite{Ade:2013zuv}); this sets our expectation for the ACT calibration to Planck.

\section{ACT and Planck data}
\label{data}

ACT is a 6-meter off-axis Gregorian telescope situated in the Atacama desert in Chile. The data used in this paper are described in D13 and consist of maps of 300 deg$^{2}$ along the celestial equator (hereafter ACT-E) and 292 deg$^{2}$ along declination -$55^{\degree}$ in the southern sky (ACT-S). We consider two seasons of data (2009, 2010) for ACT-E and one season of data (2008) for ACT-S (as these have the highest signal to noise ratio). For cross-correlating with {\it Planck}, the data within each season are divided into two splits in time, each of which is cross-linked, allowing direct estimation of the noise power spectra of the maps.
For {\it Planck} we use the `half-ring' maps from the public data release. Each frequency channel has two half-ring maps, built using only the first or the second half of the stable pointing period data. We project {\it Planck}'s galactic-coordinate HEALPix maps to patches of ACT's cylindrical equal area pixels in equatorial coordinates using the Taylor interpolation scheme described in \citet{Naess:2013uha}.

Figure \ref{Fig:MapComparison} depicts a subset of the maps used in this analysis. To account for the instrument resolution, we use the ACT beam transfer functions presented in \citet{Hasselfield:2013zza}, and for {\it Planck} we use the publicly released effective transfer functions \citep{2013arXiv1303.5068P}. The spatial variations of the {\it Planck} beam transfer functions across the sky result in effects that are significantly smaller than the statistical uncertainty in the calibration factors, and so will be ignored in this analysis. We use the ACT masks presented in D13, including a point source mask in both the ACT-S and ACT-E regions with flux cut of 15~mJy at 148 and 218 GHz, and a Galactic cirrus mask in the ACT-E region. Because the Planck beams are significantly broader than the ACT beams, point sources will appear larger in {\it Planck} maps than in ACT maps. We test the effect of widening the ACT mask by 50\% and find that our results are stable at the 0.2 $\sigma$ level.
The {\it Planck} HFI channels at 100--353 GHz are calibrated on the dipole due to the Sun's motion relative to the CMB, leading to an absolute calibration uncertainty of 0.54 percent for the 100, 143, and 217 GHz channels, with relative calibrations of 0.2 percent between them \citep{Ade:2013eta}.

\section{Cross Correlations}
\label{results}

 \begin{figure*}
\begin{center}
\includegraphics[width=17cm]{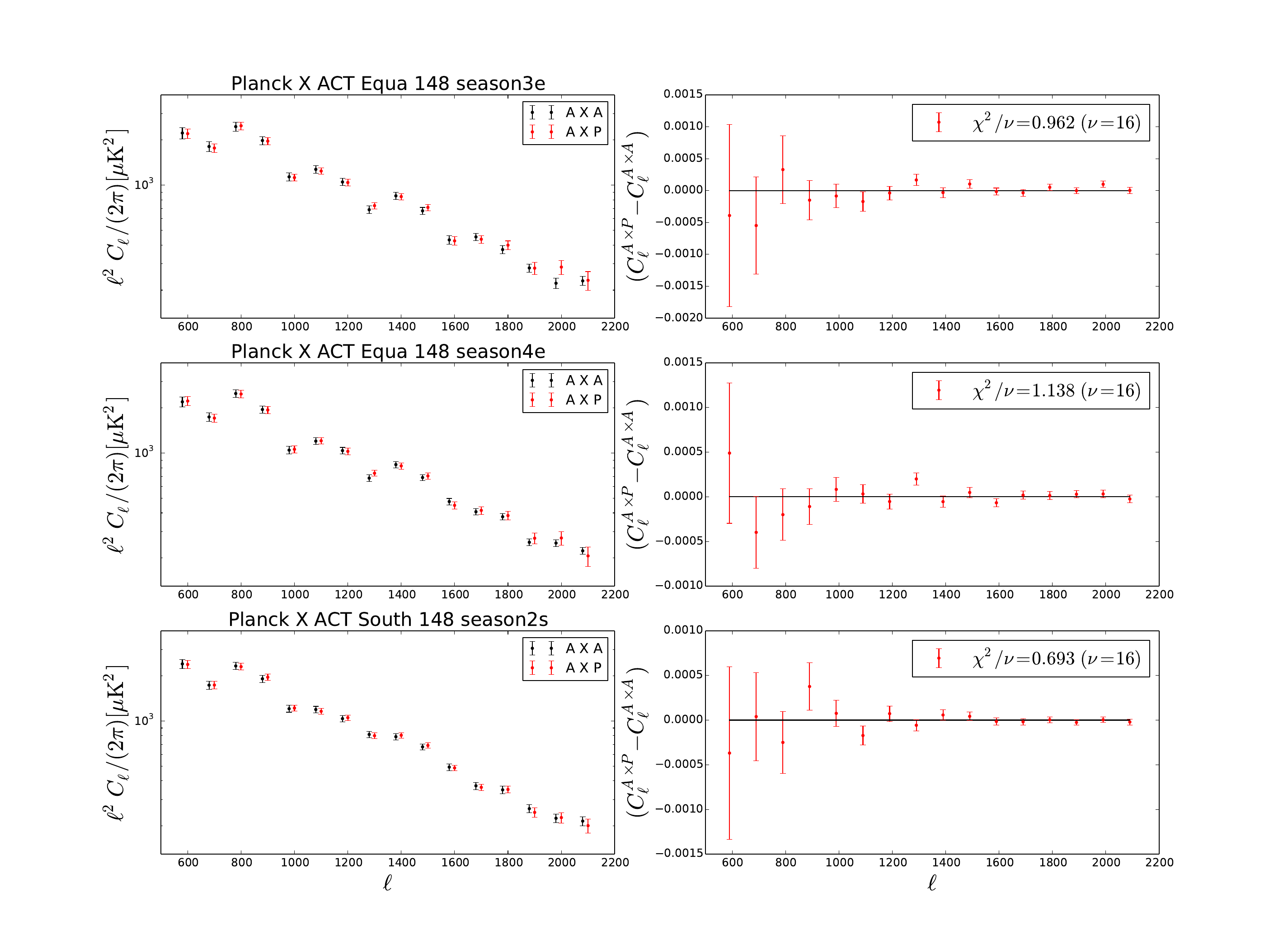}\\
\end{center}
\vskip -1cm
\caption{{\it Left:} The recalibrated cross-correlation between ACT at 148~GHz and {\it Planck} at 143~GHz (A$\times$P), compared to the recalibrated ACT power spectra (A$\times$A), in the overlapping angular range. {\it Right:} Residuals between the cross and auto-spectra as a function of scale. No significant features are observed. Since data for each experiment in each plot come from the same sky region, the errors on the residuals do not include cosmic variance.  Note also that the ordinates of the residual plots are not multiplied by the $\sim \ell^2$ factor used for plotting angular power spectra.}
\label{Fig:PlanckXACT_148}
\end{figure*}

We compute the ACT 148~GHz $\times$ {\it Planck} 143~GHz, and the ACT 218~GHz $\times$ {\it Planck} 217~GHz power spectra, for each ACT season. We follow the same procedure described in D13, including prewhitening of the maps, filtering, and deconvolving the effects of the beam, pixelization, and windowing. The uncertainties on the spectra are calculated analytically using measurements of the noise in the maps and include beam uncertainties (see Appendix \ref{apdx:errorbars}). 

\begin{figure*}
\begin{center}
\includegraphics[width=17cm]{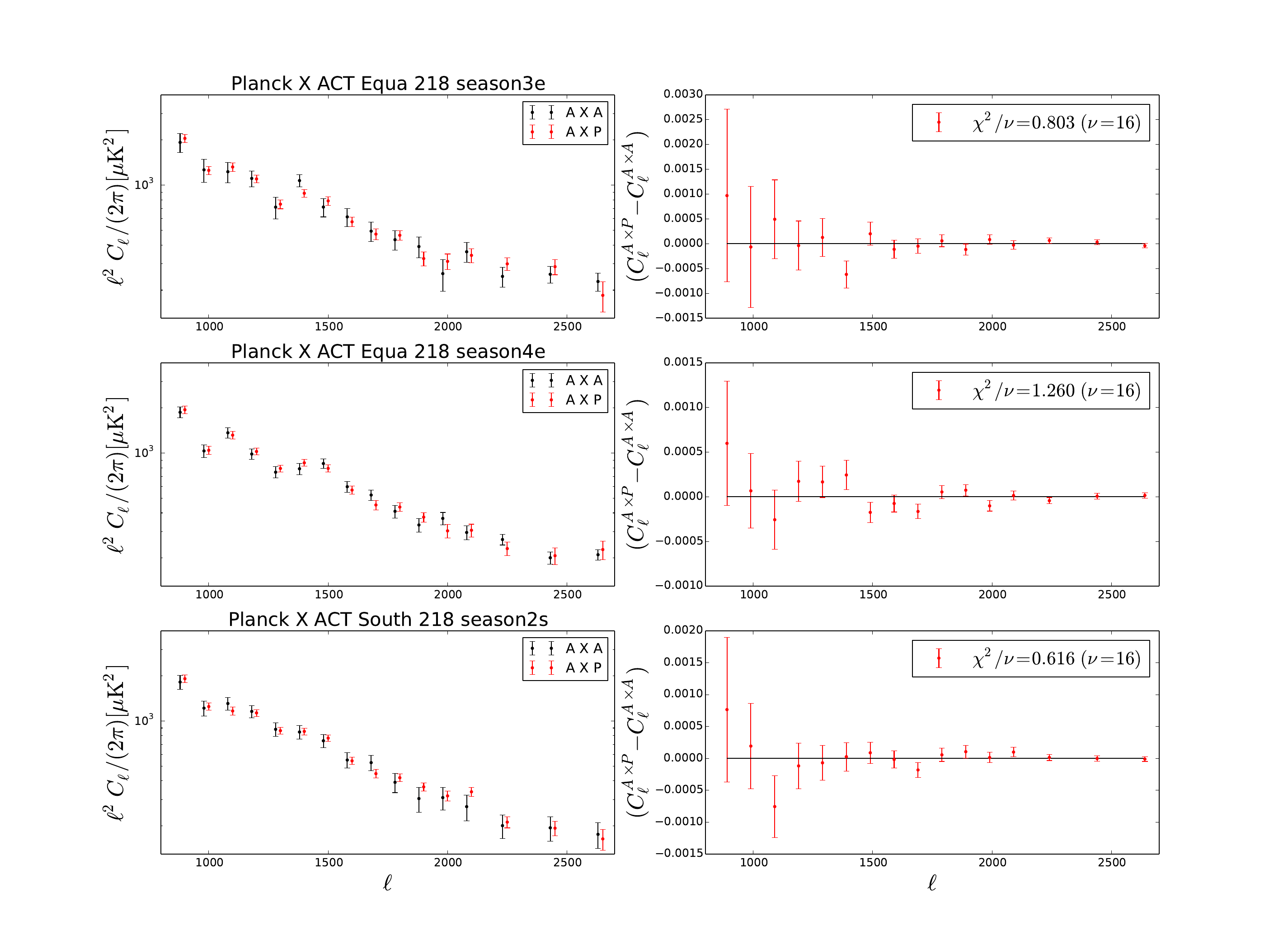}\\
\end{center}
\vskip -1cm
\caption{As in Figure \ref{Fig:PlanckXACT_148}, for the correlation between ACT at 218~GHz and {\it Planck} at 217~GHz.}
\label{Fig:PlanckXACT_220}
\end{figure*}

We allow a single degree of freedom for the comparison between the cross-spectra and the ACT spectra, namely a calibration factor that rescales the ACT power spectra to match the cross-spectra. We note that the shift in the effective frequency between the two experiments leads to a negligible variation in the foreground level. The calibration, $y$, is obtained for each frequency and ACT season and region by minimizing the $\chi^{2}$ defined as
\ba
\chi^{2}(y)= \bm{r}^{t}  \bm{\Sigma}^{-1}  \bm{r}.
\ea
Here $\bm{r}=\bm{C}^{A\times P}- y \bm{C}^{A\times A}$ is the residual between the cross and auto-spectra, which removes cosmic variance uncertainty, and $\bm{\Sigma}$ is the covariance matrix of the residual  (Appendix \ref{apdx:errorbars}). This factor $y$ is relative to the original {\it WMAP} calibration used in the D13 analysis, and can then be used to rescale the ACT maps. 
The calibrated power spectra for each season and region, together with the residuals, are shown in Figures \ref{Fig:PlanckXACT_148} and \ref{Fig:PlanckXACT_220} for the substantial range of angular scales common to both experiments.

We find the signal to be consistent at both frequencies and in both regions of the sky, with the probability to exceed (PTE) and the calibration factors reported in Table~\ref{table:cal}. The PTEs for all the spectra lie in the range $0.213<\rm{PTE}<0.874$, and there are no particular features seen in the residual spectra. We compare the ACT re-calibration factors to those determined by the {\it Planck} collaboration by jointly fitting the $\Lambda$CDM cosmological model to the ACT and full-sky {\it Planck} power spectra \citep{Ade:2013zuv}. They are reported there as $y_{148}^{\rm ACTe}, y_{148}^{\rm ACTs}, y_{218}^{\rm ACTe},y_{218}^{\rm ACTs}$, and are repeated in Table~\ref{table:cal} for comparison. The two methods give consistent results. For completeness, we also report the re-calibration factors obtained by jointly fitting the $\Lambda$CDM cosmological model to the ACT and full-sky {\it WMAP} 9 years power spectra. The difference between the {\it Planck} and {\it WMAP} calibration is consistent with our expectation.

\begin{deluxetable*}{ccllllll}
%\centering
\tablecolumns{6}
\tablewidth{0pc}
\tablecaption{Probabilities to exceed and calibration factors \label{table:cal}}

\tablehead{Freq & \colhead{ Season}Ê&$\chi^{2}$/$\nu$ &P.T.E & $y$ (this work) & $y$ (model\tablenotemark{a})  & $y_{w}$ (model\tablenotemark{b})  \\}
\startdata
%\cutinhead{{ACT 148 GHz $\times$ Planck 143 Ghz}}
%\hline
%\hline
148 & 3e & 0.962 & 0.49 &   $0.980 \pm 0.008$ &   \\
    & 4e & 1.138 & 0.31 &   $0.989 \pm 0.006$ &   \\
  &ACT-E &      &      &   &$0.988 \pm 0.007$  & $1.009 \pm 0.008$ \\
\\
&ACT-S\tablenotemark{c}
         & 0.693 & 0.804 &   $0.998 \pm 0.0065$ &  $0.992 \pm 0.007$ & $1.011 \pm 0.008$ \\

\hline 
%\cutinhead{ACT 218 GHz $\times$ Planck 217 Ghz}
%\hline
218 & 3e & 0.803 & 0.683 & $0.957 \pm 0.034$ &  \\
    & 4e & 1.26 & 0.213 & $0.969 \pm 0.02$ & \\
     &ACT-E &  &  & & $0.96 \pm 0.01$ & $0.99 \pm 0.01$  \\
\\ 
  &ACT-S & 0.616 & 0.874  & $1.001 \pm 0.025$ & $1.01 \pm 0.01$  & $1.04 \pm 0.02$ 
\enddata
\footnotetext[1]{From \citet{Ade:2013zuv}.}
\footnotetext[2]{ Re-calibration factors obtained by jointly fitting the $\Lambda$CDM cosmological model to the ACT and full-sky {\it WMAP} 9 years power spectra.}
\footnotetext[3]{ACT-S includes just the correlation with the 2008 maps.}
\label{table:cal}
\end{deluxetable*}

 \section{Galactic dust and CIB using {\it Planck} 353~Ghz} 
\label{dust}
\begin{figure}[]
\begin{center}
\includegraphics[width=9.5cm]{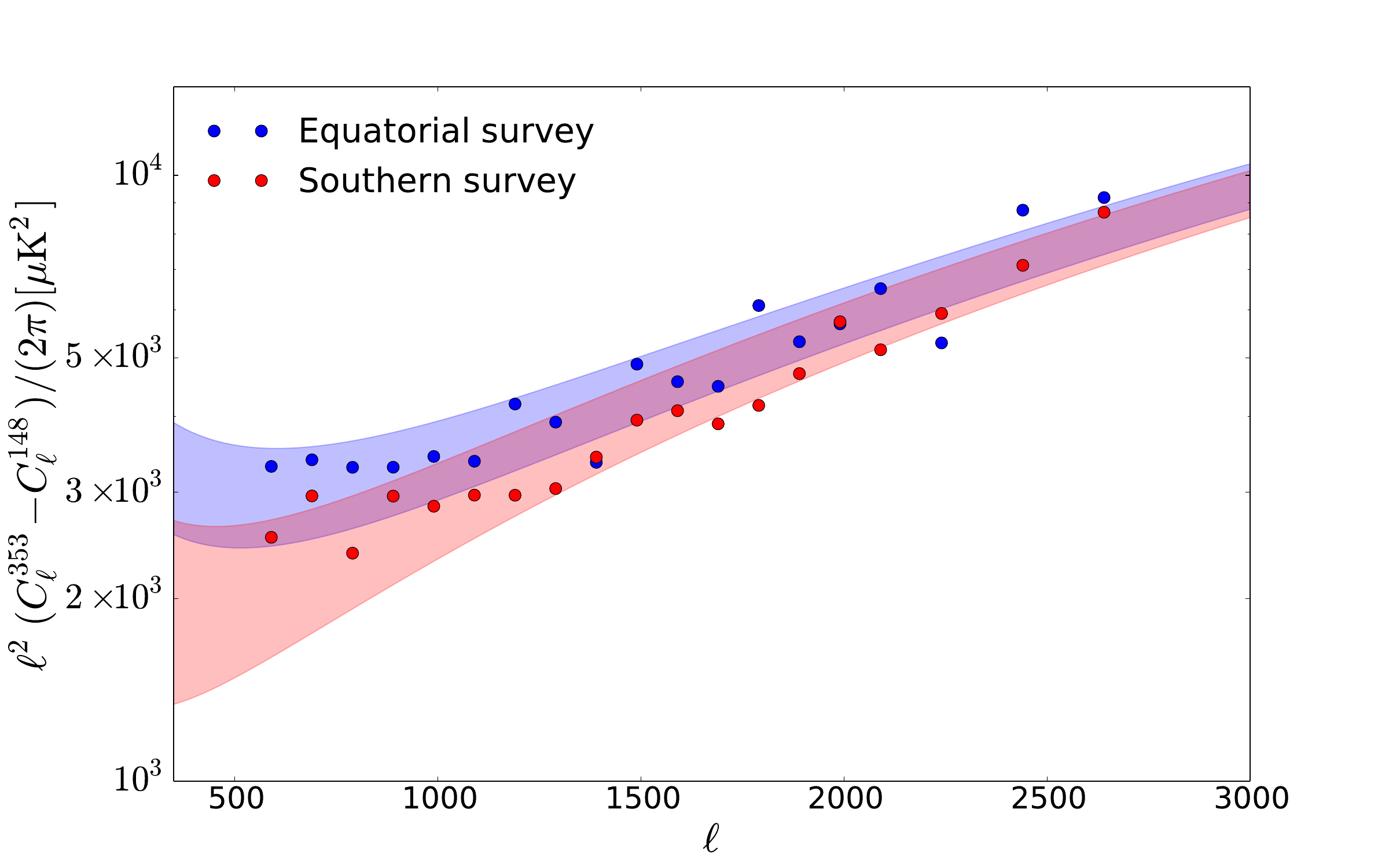}\\
\end{center}
\caption{Difference between the measured \emph{Planck} 353~GHz and ACT 148 GHz power spectra for the equatorial and southern surveys (dots). The colored bands represent the $1\sigma$  prediction for the Galactic dust and CIB amplitude based on the extrapolation of the ACT foreground power spectra modeling at 148 and 218 GHz  \citep{Dunkley:2013vu}.}
\label{Fig:dust}
\end{figure}

The 353 GHz {\it Planck} maps can be modeled as the sum of CMB, Galactic cirrus dust and CIB fluctuations. 
Figure \ref{Fig:dust} shows the difference between the measured {\it Planck} 353~GHz and ACT 148 GHz power spectra for the equatorial and southern surveys.
Since the dust emission is inhomogeneous, we use the ACT spatial weighting scheme when computing the {\it Planck} 353 GHz power spectrum. 
The difference between the two power spectra is dominated by Galactic cirrus and CIB fluctuation emissions at 353 GHz.
The contribution from other signals, such as the Sunyaev Zel'dovich effect, is subdominant and can be safely neglected.

Figure \ref{Fig:dust} also shows predictions and error bands for the sum of the CIB and cirrus contributions at 353 GHz. We estimate the Poisson and clustered CIB contribution at 353~GHz based on the ACT constraints from Table 2 of \cite{Dunkley:2013vu}. 
 
 Following \cite{Addison:2011se}, the CIB power spectrum  is modeled as the square of a modified blackbody in flux density units with emissivity index constrained to be $\beta = 2.2 \pm 0.1$, with a fixed effective temperature of $T = 9.7$ K. As can be seen in Figure 17 of the \cite{2013arXiv1309.0382P} paper,  the 2013 Planck CIB model is $\sim 30\%$ higher than the early Planck 217 GHz CIB bandpowers that were used to constrain the Addison et al. (2011) model.  We renormalize the Addison et al. model  by multiplying  the 353 GHz prediction by a factor of 1/1.3.

The Galactic dust power at 353~GHz was calculated using a modified blackbody with emissivity index $\beta=1.5$ and temperature $T=20~$K, consistent with submillimeter analysis of cirrus in a similar region of sky along the celestial equator \citep{bracco/etal:2011}, and with observations from the Planck satellite  \citep{Abergel:2013fza}.  We correct the Galactic and CIB power calculated at the nominal frequency, 353~GHz, for the bandpass profile and response to a dust-like source SED \citep{2013arXiv1303.5070P}. The uncertainties shown in Figure \ref{Fig:dust} are dominated by uncertainty in the CIB and Galactic dust amplitudes measured by ACT \citep{Dunkley:2013vu}.
We interpret the figure as showing that the ACT dust model and measurements are in agreement with {\it Planck}'s measurement at 353 GHz.

\section{Isotropy of the two-dimensional power spectra}
\label{Iso}

\begin{figure*}
\centering
\includegraphics[width=16.5cm]{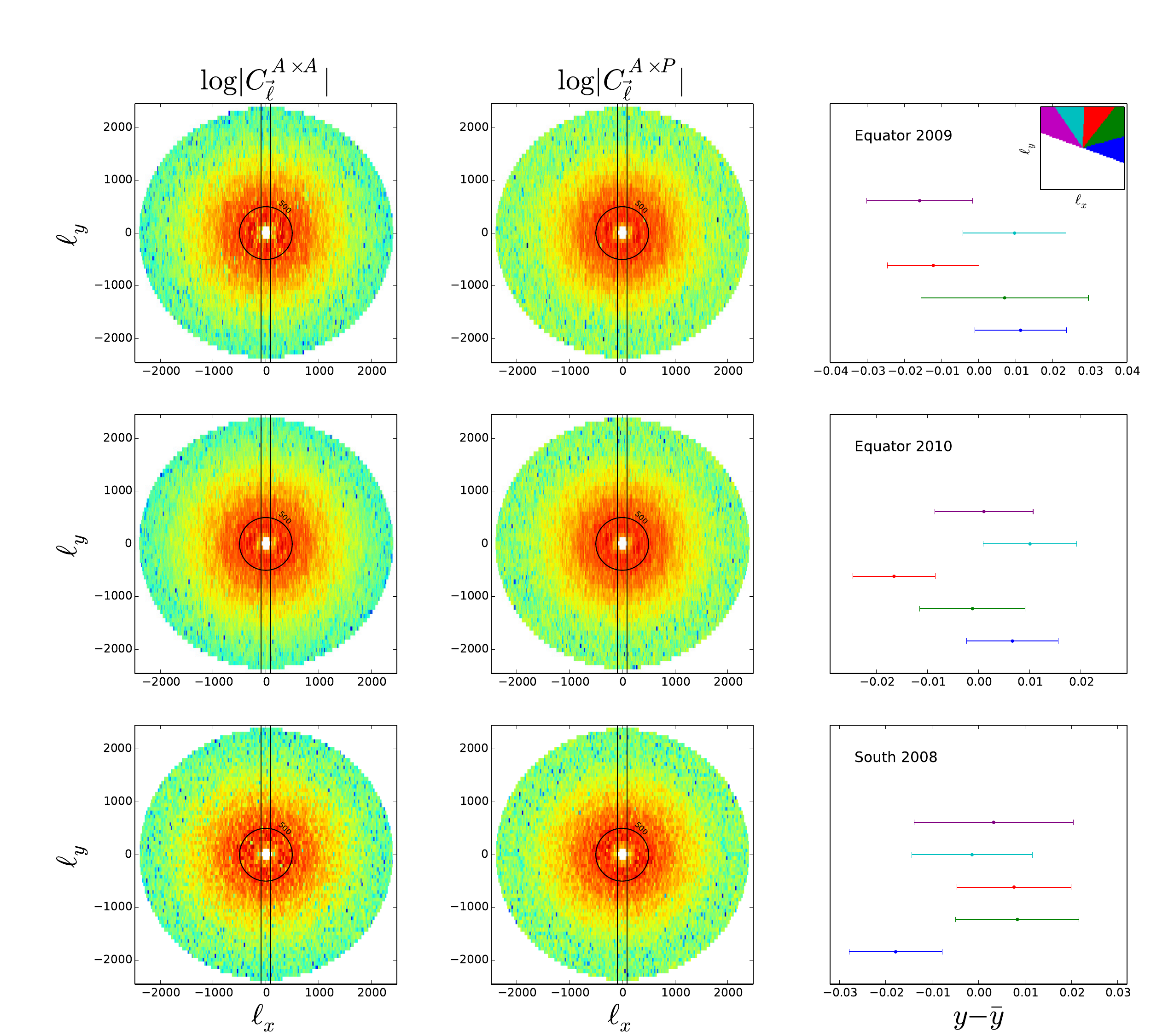}\\
\caption{ Two-dimensional ACT$\times$ACT power spectra (left) and  ACT$\times${\it Planck} (middle)  in the overlapping angular range. The vertical bands show the fourier mask applied to the ACT spectra to avoid artifacts of the scanning strategy and the black circles encompass the $\ell<500$ modes that are not used in the power spectra analysis. {\it Right:} calibrations as a function of the angular bands, from which we have subtracted their mean. No significant anisotropy is detected. }
\label{Fig:2dComparisonAR1}
\end{figure*}

\begin{figure*}
\centering
\includegraphics[width=16.5cm]{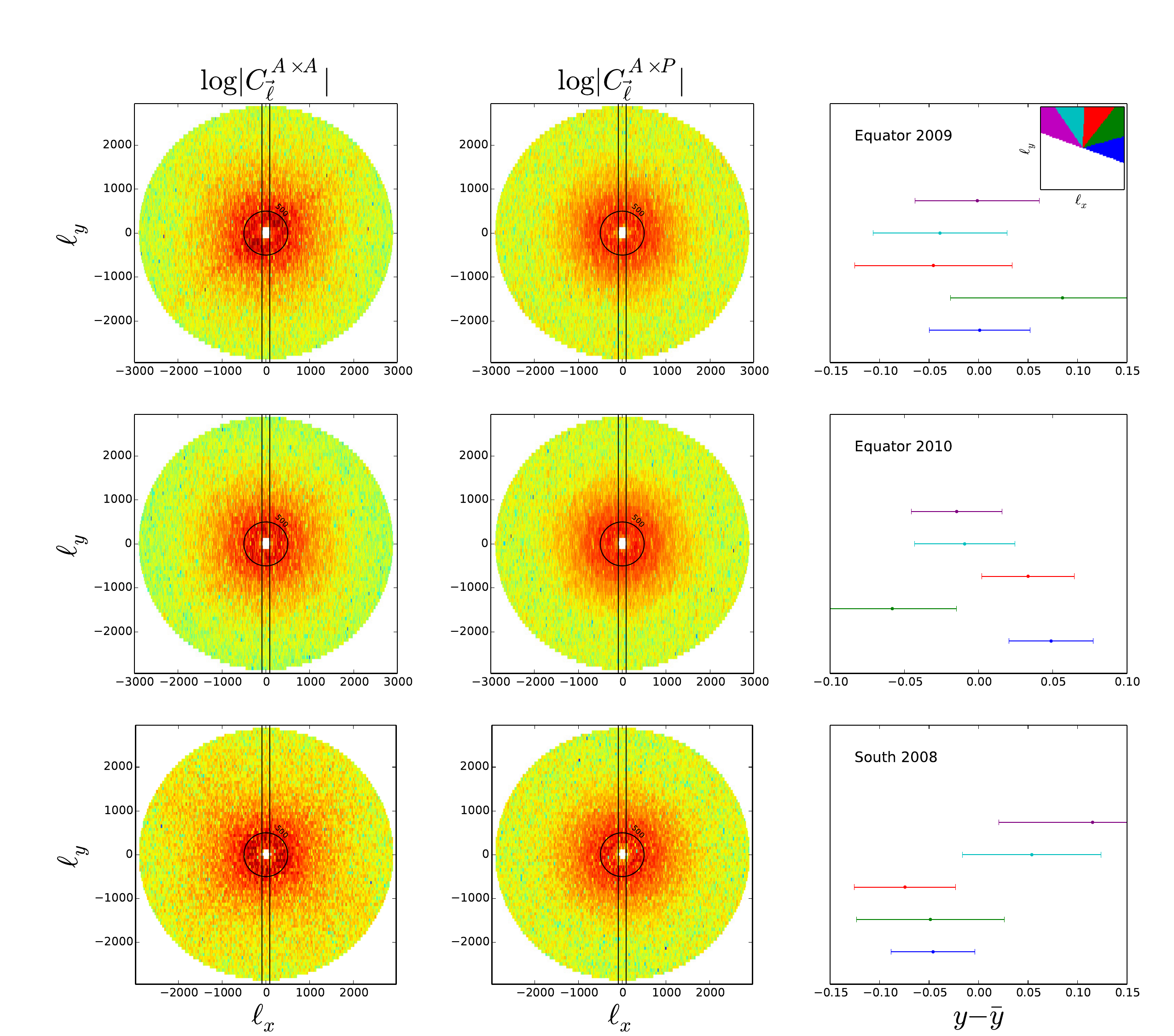}\\
\caption{ As in Figure \ref{Fig:2dComparisonAR1}, for the correlation between ACT at 218~GHz and {\it Planck} at 217~GHz. Here the absolute value emphasize the ÒXÓ shaped patterns.}
\label{Fig:2dComparisonAR2}
\end{figure*}

Because of its scan strategy, ACT has anisotropic noise.  While the map-making algorithm should not produce
any anisotropy in the map or the transfer function, the {\it Planck} data provide an opportunity to test the ACT maps
by checking for any directional dependence in the calibration factor.
Excess noise along and perpendicular to the ACT scan directions leads to ÒXÓ shaped patterns of high noise regions in the ACT two-dimensional power spectra.
These features have been down-weighted accordingly using azimuthal weights in the D13 analysis. {\it Planck} data do not suffer from such artifacts and the effects due to the scanning strategy of ACT become subdominant in the cross-spectra. We compare the two-dimensional power spectra of ACT and ACT $\times$ {\it Planck} in order to assess the isotropy of the deduced calibration parameter. 
We compute the two-dimensional power spectra following the same procedure as for the one-dimensional analysis, except that we do not deconvolve the effect of the window function, because the inversion of the un-binned mode coupling matrices is too computationally intensive. We divide the two-dimensional power spectrum into five different angular wedges, of 36$^{\degree}$ each, using the symmetry $\vec{\ell} \rightarrow  \vec{-\ell}$ of the power spectra.
We then bin the power spectra using the same bin size as for the one-dimensional power spectra and compute the expected variance in each bin using Monte-Carlo simulations. Finally, we compute the best-fit calibration number for each angular bin. We do not detect any significant anisotropy in the two-dimensional power spectra.
The best-fit calibration per angular bin as well as the two-dimensional power spectra are shown in Figure \ref{Fig:2dComparisonAR1} and Figure \ref{Fig:2dComparisonAR2}.

\newcommand{\cleft}[0]{14mm}
\newcommand{\cright}[0]{14mm}
\newcommand{\ctop}[0]{0mm}
\newcommand{\cbot}[0]{0mm}

 \section{Planet Brightness Temperatures} 
\label{planet}

The calibration correction factors may be used to refine
the ACT planetary brightness temperature measurements originally
presented in \citet{Hasselfield:2013zza}.  The ACT results may then be
compared to {\it Planck} measurements to demonstrate the extent to which the
ACT beam, which transfers the instrument calibration to small angular
scales, is understood.

Using the factors presented in Table 1, we recalibrate the ACT
measurements of Uranus for each season and array, obtaining mean
Rayleigh-Jeans temperatures of $102.2 \pm 1.7$~K at 149~GHz and $92.5
\pm 2.8$~K at 219~GHz.  The errors are now dominated by uncertainties
in the spectral bandpass rather than in the calibration.  Our
measurements of Uranus are within 1$\sigma$ of the temperatures
measured by \cite{Ade:2013eta} at 143~GHz and 217~GHz, respectively.

The revised coefficients for the empirical model of Uranus brightness
temperature presented in \citet{Hasselfield:2013zza} are found to be
$(a_0, a_1, a_2)$ = (120, -80.9, 22.2), which yields temperatures in
the 100 to 300~GHz range that are up to 1.6\% lower.  The resulting
model lies within the 1$\sigma$ uncertainties of the {\it Planck}
measurements at 100, 143, 217 and 353~GHz, though those data were not
used to constrain the model.

We also recalibrate the ACT Saturn brightness temperature measurements, and refit both the disk+ring and disk-only models.  The fit quality and uncertainties change only slightly, and the inferred Saturn disk temperatures decrease by roughly 1\% at 149 GHz and 2\% at 219 GHz.

The cross-correlation calibration is most sensitive to angular scales
in the range of $\ell =500$ to 2000, while ACT's flux density
measurements of compact sources such as Uranus are sensitive to much
finer angular scales, around $\ell = 8000$.  The agreement
between {\it Planck} and ACT brightness temperatures demonstrates that the
ACT beam and its solid angle have been characterized to better
than  3\% (with this number coming from the {\it Planck} planet
brightness temperature uncertainties).

 \section{Sources} 
 \label{sources}
A separate but related method of determining the calibration factor $y$ is to compare ACT and {\it Planck} measurements of unresolved sources.  Determining the flux density of a compact source requires both a calibrated map and knowledge of the relevant beam solid angle.  Comparing {\it Planck} and ACT measurements can thus test our understanding of both instruments' beams. 

Nearly 60 of the ACT sources tabulated in \cite{2013arXiv1306.2288M} and Gralla et al. (2014, in preparation) were detected in one or both of {\it Planck}'s 143 and 217 GHz bands.  The detected sources are among the brightest in the ACT lists, and all but two of them are classified as synchrotron emitters.  Most are blazars \citep{Aatrokoski:2011bg}, a class of source known to be variable on time scales of weeks to years.  The ACT observations were not necessarily made at the same time as the {\it Planck} observations (obviously the case for the 2008 ACT runs, made before  {\it Planck} began observations in August 2009).  We expect source variability to introduce scatter in plots comparing Planck and ACT flux densities.  We study this issue further in Section 7.4 below.

\subsection{ACT data}

The 2008 ACT South flux densities are from \cite{2013arXiv1306.2288M}.  A few additional southern sources that fall outside the sky area treated in that paper are also included.  Flux densities for equatorial sources are from Gralla et al. (2014, in preparation).  For the equatorial sources, which were observed in two seasons, we made inverse-variance weighted averages of the 2009 and 2010 values.

\vspace{1.5cm}
\subsection{Planck data}

The {\it Planck} measurements used here are taken from the {\it Planck} Catalogue of Compact Sources (PCCS)\footnote{Available at  http://irsa.ipac.caltech.edu/Missions/Planck.html}; see \cite{2013arXiv1303.5088P}.  {\it Planck} measurements at 143 and 217 GHz were used.  We employed DETFLUX values, as described in the PCCS, in order to reduce sensitivity to neighboring sources, extended structure and background emission that could enter {\it Planck}'s broader beams. PCCS flux densities for a given source are averages of all observations of that source made during the period 12 August 2009 to 27 November 2010.  These flux densities require small color corrections of a few per cent, since the sources treated here have very different spectra from the CMB dipole used to calibrate the {\it Planck} maps.  These color corrections depend on the spectral index of each source.  Each index was computed from the {\it Planck} data, and was also used to make the small extrapolations from {\it Planck} band centers at 143 GHz and 217 GHz to the corresponding ACT band centers appropriate for synchrotron sources \citep{Swetz:2010fy}, 147.6 and 217.6 GHz.  For a handful of weak sources, {\it Planck} data were missing at one of the two frequencies; we assumed typical values of the spectral index for these sources.  At 218 GHz, the color correction is partially canceled by the small extrapolation in frequency, so the resulting multiplicative correction to the PCCS values varied only between 0.984 and 0.997.  At 148 GHz, the correction is larger, ranging from 0.96-1.02 for all but the two inverted spectrum dusty sources.

\subsection{Comparing Planck and ACT flux densities}

\begin{figure}[]
\begin{center}
\includegraphics[width=9cm]{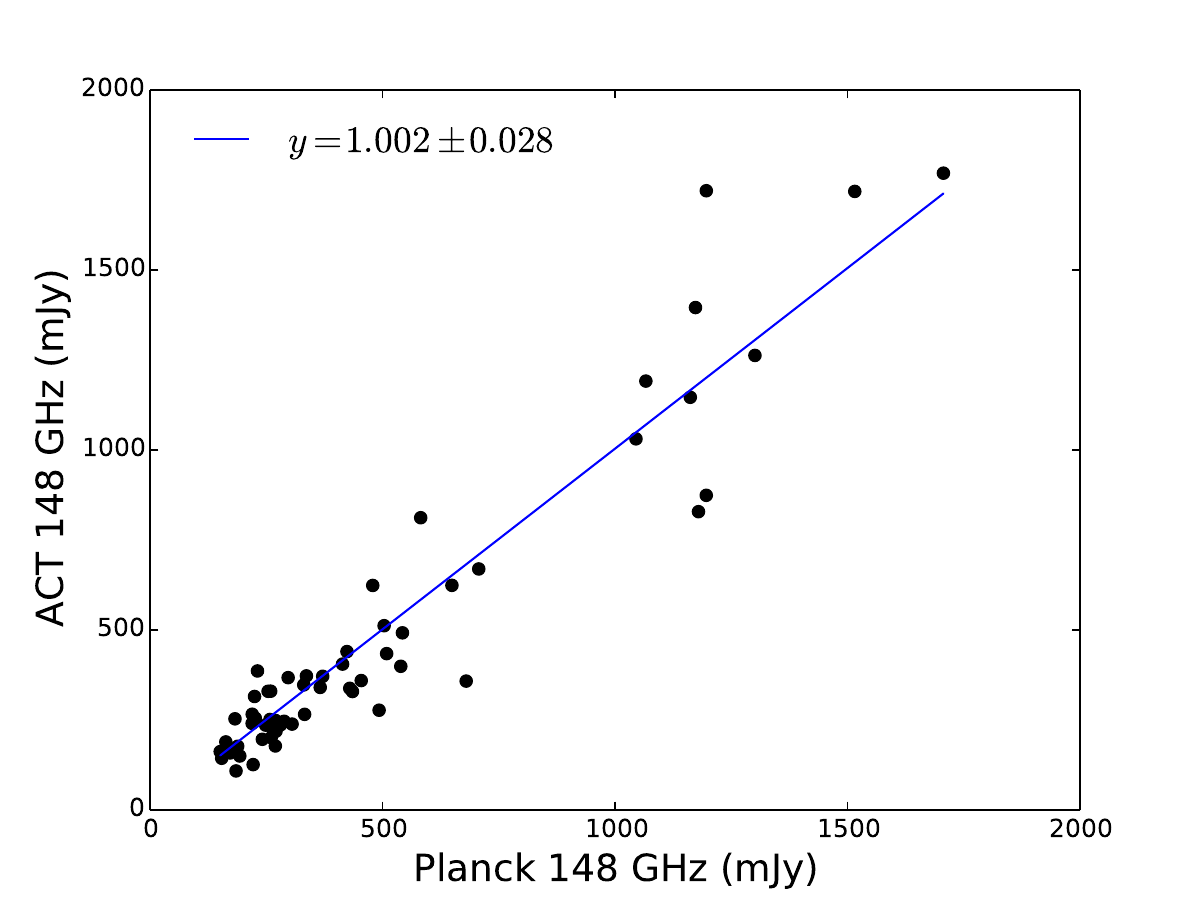}\\
\end{center}
\caption{Comparison of ACT (vertical axis) and {\it Planck} (horizontal axis) flux densities at 148 GHz. The calibration factor $y$ is consistent with the one obtained using the power spectra.}
\label{Fig:SourcesAR1}
\end{figure}

\begin{figure}[]
\begin{center}
\includegraphics[width=9cm]{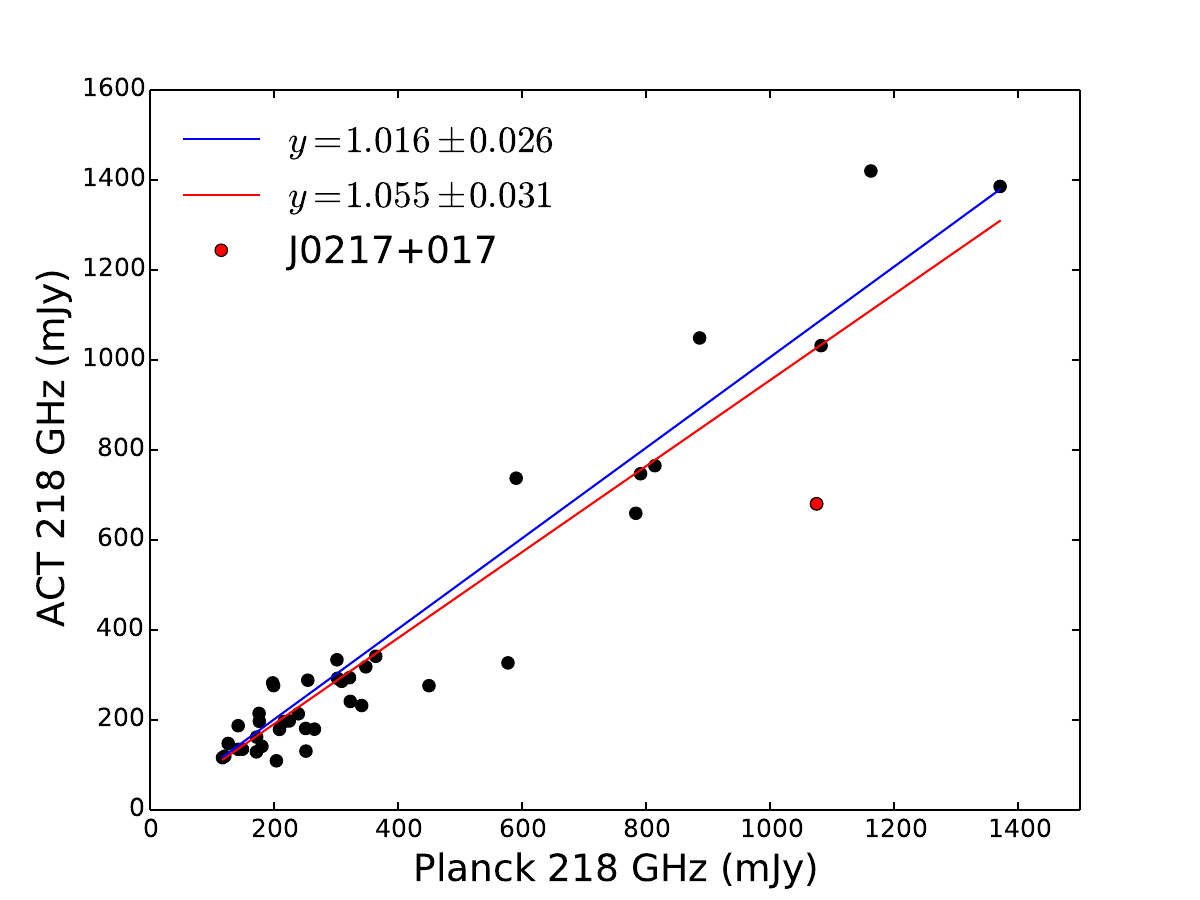}\\
\end{center}
\caption{As in Figure \ref{Fig:SourcesAR1}, for  ACT and {\it Planck} at 218~GHz. We also show the effect of dropping the variable source J0217+017.}
\label{Fig:SourcesAR2}
\end{figure}

Figure \ref{Fig:SourcesAR1} and \ref{Fig:SourcesAR2} compare the ACT and {\it Planck} flux densities. For both frequencies, the linear fits were forced to pass through the origin (see Planck Collaboration XIV, 2011) to reduce the effect of source boosting, sometimes called Eddington bias, in the {\it Planck} data.  Relaxing this constraint typically changes the slope of the fits by less than $1 \sigma$. At 148 GHz, the agreement is excellent, despite the evident scatter introduced by variability. ACT's flux densities agree with {\it Planck}'s to within 1$\%$; from this comparison we find $y=1.002 \pm 0.028$. The uncertainties in flux density in the {\it Planck} measurements, at 30 to 40 mJy, are typically 10 times those in the ACT measurements: the size of the symbols used in the figures is roughly equivalent to the $1 \sigma$ {\it Planck} errors.

At 218 GHz, there are fewer (45) matched sources (and the source variability may be greater).  At 218 GHz, ACTÕs flux densities run about 5\% lower than 
{\it Planck}'s; formally, with all the data included, the implied correction factor $y = 1.055 \pm 0.031$.  This value is higher ($\sim 2 \sigma$) than the value of $y$ obtained from the power-spectrum comparisons (see Table 1). We explore the issue of variability in Section \ref{variability}  below.  A second possibility for this apparent discrepancy is that {\it Planck}'s larger beam is picking up emission from sources clustered around the primary source.  Such an effect has been seen in the case of lensed, submillimeter sources by Welikala et al (2014, in preparation).  These authors ascribe excess {\it Planck} flux to sources clustered around a lensing source, however,  and there is no evidence that the blazars that make up the vast majority of the sources used here are lensed.

\subsection{The effect of source variability}
 \label{variability}
So far, we have implicitly assumed that the expected variability of the synchrotron sources studied here does not, on average, bias the ACT-{\it Planck} comparison.  We can test that assumption in two different ways.  First, we can arbitrarily exclude discrepant points from the graphs to see what effect there is on the fitted slopes.  A second, and better justified, approach is to drop sources for which we have evidence of variability.  The latter is possible for the ACT equatorial sources, many of which were observed at two epochs roughly a year apart. Two such variable sources are J0217+017 and J0739+016.
At 148 GHz, dropping 4 or 5 of the most divergent points from Figure \ref{Fig:SourcesAR1} changes the slope of the fit, and hence $y$, by $\sim$ 2$\%$, or $0.8 \sigma$.  The same is true for dropping 4 to 5 of the sources observed to be variable, such as J0217+017.  In either case, a value of $y$ = 0.98-0.99  is favored, in agreement with the results of section 3. 

At 218 GHz, in contrast, dropping variable sources makes a larger difference.  If we drop just one known variable source, J0217+017, $y$ changes from 1.055 to 1.016, a change of $\sim 1.4 \sigma$.  If  we drop 3-4 more variable sources, $y$ remains at 1.01 $\pm$ 0.026, with no significant change in the associated statistical error.  This value is now more consistent with the results from Section 3.  

\subsection{Other tests}
\label{othest}

We conducted several tests of the stability of these results on compact sources. First, as noted, allowing an unconstrained fit to the data changed the slope of the fits (and hence $y$) by less than $1\sigma$. Another test of the effect of possible Eddington bias in the { \it Planck} data was to drop the weakest sources (around 20$\%$ of the total). There was little effect on the values of the calibration factor $y$. We also repeated the fits with unweighted averages of the 2009 and 2010 ACT data, rather than inverse-variance weighted averages. At 148 GHz, using unweighted averages raised $y$ by 2$\%$ or less than 1$\sigma$. At 218 GHz, with J0217+017 omitted, the value of $y$ shifted only slightly, from 1.016 to 1.020.
Next, we omitted all the 2008 ACT data, taken before {\it Planck}Õs launch. At 218 GHz, $y$ changed minimally from 1.016 to 1.004 (again excluding J0217+017). The same test at 148 GHz resulted in a small change in $y$, from 0.983 to $0.971 \pm 0.017$.
Finally, we tried dropping $\sim$5 -10 sources at low Galactic latitude or those flagged in the PCCS as possibly contaminated by Galactic cirrus emission (with the ÒCIRRUSÓ flag $>$ 10;  see Planck Collaboration XXVIII, 2013). Two of the 148 GHz sources dropped were known variables, J0253-544 and J0739+016.
The result was to lower $y$ by $\sim 2 \%$ at 148 GHz, with a much smaller effect at 218 GHz.
None of these tests resulted in a change of $y$ greater than 1$\sigma$.

 \section{Conclusion} 
\label{conclude}

The ACT experiment has mapped two regions of the sky covering 592 deg$^2$ at arcminute resolution. The same regions have now been observed by the {\it Planck}  satellite, and we have computed the cross-correlation to check for consistency between the data sets. The cross-comparison also tests the beams and transfer functions of both experiments. The cross-power is consistent at all angular scales probed by both experiments, and at both frequencies observed by ACT. We have estimated new calibration factors for ACT at higher precision by using the {\it Planck} data in place of {\it WMAP}.  The isotropy of the calibration factor implies that the ACT scan strategy did not introduce anisotropies into the maps. We have also measured the {\it Planck} 353 GHz power spectrum in the ACT equatorial and southern surveys region, and found the amplitudes of the dust and CIB to be consistent with those measured by ACT. We have used the new calibration factors to rescale the ACT planetary brightness temperature measurements and found them to be consistent with {\it Planck}. ACT and {\it Planck} measurements of compact sources provide results consistent with those found from a comparison of power spectra, albeit at lower sensitivity and with an extra uncertainty associated with source variability. Overall the agreement between the two measurements is excellent.

%%%%%%%%%%%%%%%%%%%%%%%%%%%%%%%%%%%%%%%%%%%%%%%%%%%%%%%%%%%%%%%%
\acknowledgements

This work was supported by the U.S. National Science Foundation through awards AST-0408698 and AST-0965625 for the ACT project, as well as awards PHY-0855887 and PHY-1214379. Funding was also provided by Princeton University, the University of Pennsylvania, and a Canada Foundation for Innovation (CFI) award to UBC. ACT operates in the Parque Astron\'omico Atacama in northern Chile under the auspices of the Comisi\'on Nacional de Investigaci\'on Cient\'ifica y Tecnol\'ogica de Chile (CONICYT). Computations were performed on the GPC supercomputer at the SciNet HPC Consortium. SciNet is funded by the CFI under the auspices of Compute Canada, the Government of Ontario, the Ontario Research Fund -- Research Excellence; and the University of Toronto. Funding from ERC grant 259505 supports JD, EC, SN and TL. We thank George Efstathiou and Duncan Hanson for discussions about the Planck data. We acknowledge the use of the Legacy Archive for Microwave Background Data Analysis (LAMBDA). Support for LAMBDA is provided by the NASA Office of Space Science.

 \bibliography{act}

\appendix
\section{Analytic error bars}
\label{apdx:errorbars}

The total covariance matrix  $\bm{\Sigma}$ of the residual $C_{b}^{A\times A}-C_{b}^{A\times P}$  is a sum of two terms. The first accounts for the noise in ACT and {\it Planck} and the second accounts for the beam uncertainties:  $\bm{\Sigma}=\bm{\Sigma}_{n} + \bm{\Sigma}_{\rm beam} $. The noise term is given by
\ba
\bm{\Sigma}_{n} = \left\langle ( C_{b}^{A\times A}-C_{b}^{A\times P}) ( C_{b}^{A \times A}-C_{b}^{A\times P})  \right\rangle &=& \Theta^{Ê (A\times A); ( A\times  A)}_{b}+\Theta^{Ê (A\times P); ( A\times  P)}_{b}-2 \Theta^{Ê (A\times A); ( A\times  P)}_{b}.
\ea
Each of the terms can be computed analytically:
\ba
\Theta^{Ê( A\times A); ( A\times  A)}_{b} &=& \frac{1}{ \nu_{b}} \left[ 2 C_{b}^{2} +4 \frac{C_{b}}{n_{d}}N_{b}^{AA}+2 \frac{(N_{b}^{ AA})^{2}}{n_{d}(n_{d}-1)} \right] \\
\Theta^{Ê( A\times  P);Ê( A\times P)}_{b} &=& \frac{1}{ \nu_{b}} \left[ 2 C_{b}^{2}+ \frac{C_{b}}{n_{d}}( N_{b}^{ AA}+N_{b}^{ PP})+\frac{N_{b}^{AA}N_{b}^{ PP}}{n_{d}^{2}}\right] \\
\Theta^{Ê( A\times  A);Ê( A\times  P)}_{b}&=&  \frac{1}{\nu_b} \left[ 2C_{b}^{2} + 2\frac{C_{b}N_{b}^{AA}}{n_{d}} \right], 
\ea
where $n_{d}$ is the number of data splits and $\nu_{b}$ is the number of modes per bin, corrected for the effect of the window function.  $C_{b}$ is a theoretical power spectrum, and $N_{b}^{XX}$ is the noise power spectrum, given by $C_{b,\,{\rm auto}}^{XX} -C_{b,\,{\rm cross}}^{XX} $.
Finally the full covariance is given by
\ba
 \left\langle ( C_{b}^{A\times A}-C_{b}^{A\times P}) ( C_{b}^{A \times A}-C_{b}^{A\times P}) \right\rangle=  \frac{1}{ \nu_{b}} \left[ 2 \frac{(N_{b}^{ AA})^{2}}{n_{d}(n_{d}-1)} + \frac{C_{b}}{n_{d}} (N_{b}^{ PP}+N_{b}^{ AA}) +\frac{N_{b}^{AA}N_{b}^{ PP}}{n_{d}^{2}}\right].
\ea

\end{document}